\documentclass[12pt]{iopart}

\usepackage{iopams}
\usepackage{graphicx}
\usepackage{amssymb, amsthm}%\usepackage{amsmath}
\usepackage[draft]{hyperref} %% optional
\usepackage{setspace} %% setting single or double space

\begin{document}

\title{Harmonic oscillations and their switching in elliptical optical waveguide arrays}

\author{Ming Jie Zheng,$^{1}$ Yun San Chan,$^1$ and Kin Wah Yu$^{1,2}$}
\address{$^1$Department of Physics, The Chinese University of Hong Kong, Shatin, New Territories, Hong Kong, China}
\address{$^2$Institute of Theoretical Physics, The Chinese University of Hong Kong, Shatin, New Territories, Hong Kong, China}
\ead{mjzheng@phy.cuhk.edu.hk}

\begin{abstract}
We have studied harmonic oscillations in an elliptical optical waveguide array in which the coupling between neighboring waveguides is varied in accord with a Kac matrix so that the propagation constant eigenvalues can take equally spaced values. As a result, long-living Bloch oscillation (BO) and dipole oscillation (DO) are obtained when a linear gradient in the propagation constant is applied. Moreover, we achieve a switching from DO to BO or vice versa by ramping up the gradient profile. The various optical oscillations as well as their switching are investigated by field evolution analysis and confirmed by Hamiltonian optics. The equally spaced eigenvalues in the propagation constant allow viable applications in transmitting images, switching and routing of optical signals.

\hfill


\end{abstract}

%\ocis{130.2790, 130.4815, 230.4910, 230.7370}

\maketitle

\newpage

\section{Introduction}

Optical oscillations in optical waveguide arrays (OWA) are harmonic when the propagation constants of the transverse modes are equally spaced \cite{PhysRevA.63.063410.2001, IEEE.49.745.2004, OptLett.29.2752.2004}. These oscillator waveguide arrays are of particular interest not only due to their applications in transmitting images \cite{IEEE.71.208.1983}, focusing and steering \cite{PhysRevLett.103.033902.2009}, tunneling \cite{PhysRevLett.96.023901.2006, PhysRevLett.96.053903.2006, PhysRevLett.102.076802.2009}, and switching and routing optical signals \cite{PhysRevLett.83.4752.1999, PhysRevLett.83.4756.1999}, but they are also good candidates for realizing the optical analogies of the dynamics in quantum systems \cite{LaserPhoton.Rev.3.243.2009}. Among these, Bloch oscillation (BO) and dipole oscillation (DO) are two fundamental and important types of oscillations \cite{Nature.424.817.2003, PhysRep.463.1.2008}. BO is known as the oscillatory motion of a particle in a periodic potential under a constant force \cite{NewJPhys.6.2.2004}, whose optical equivalent in an OWA is a linear gradient of the onsite propagation constants \cite{OptLett.23.1701.1998, PhysRevLett91.263902.2003}. DO usually occurs at the bottom of a parabolic band \cite{NewJPhys.5.112.2003}. In the reciprocal space, DO distinguishes from BO that the BO momentum is accelerated monotonically by the force while the DO momentum oscillates due to the confinement of a band \cite{PhysRevA.81.033829, JOptSocAmB.27.1299.2010}. The steering between BO and DO has been realized in a parabolic optical waveguide array \cite{JOptSocAmB.27.1299.2010}. However, both BO and DO decay fairly quickly due to the unequally spaced propagation constants in a parabolic band. In order to obtain long-living BO and DO, we need to construct OWAs so that the eigenvalues of the propagation constants are equally spaced.

In the literature, several models have been proposed to achieve this goal, for example, the Wannier-Stark waveguide arrays, which require an infinitely varying transverse refractive-index distribution \cite{OptLett.23.1701.1998}. In real experiments \cite{PhysRevLett.83.4752.1999, PhysRevLett.83.4756.1999}, we need to truncate the system, which unavoidably disrupt the equally spaced propagation constants. As a result, the oscillations can decay rapidly, especially when the light beam is incident at the boundaries of the system. Besides, there also exist dissipation \cite{OptLett.29.2485.2004}, defects \cite{PhysRevB.81.195118.2010}, and other kinds of damping effects in real systems. Recently, a finite waveguide array with graded coupling constants described by the elements of a Kac matrix was proposed \cite{Algebr.150.341.1991, Proceedings.503.1994}. This is to achieve the required equally spaced propagation constants \cite{OptLett.29.2752.2004}. From now on, we refer these graded couplings $\kappa$ to circular couplings as $\kappa$ varies circularly as a function of waveguide index. In the finite waveguide arrays with a constant on-site potential, harmonic oscillation other than BO has been observed. The harmonic oscillation is similar to DO, as it also oscillates in the reciprocal space.

In this work, we apply a linear gradient to the onsite propagation constants, which leads to the occurrence of BO. Strictly speaking BO is approximate, because the band structure is elliptical caused by the combination of linear graded onsite potential and the circular couplings. Due to the elliptical profile of the dispersion relation, we call the array as the elliptical optical waveguide array (EOWA). We propose to realize a switching between DO and BO or vice versa by ramping up the gradient of propagation constants. The DO-BO transition is investigated through the field-evolution analysis and confirmed by the Hamiltonian optics approach. The equally spaced propagation constants and switching between DO and BO have potential applications in optical steering devices.

\section{Model and formalism}
\begin{figure}[htbp]
  \centering
  \includegraphics[width=0.5 \textwidth]{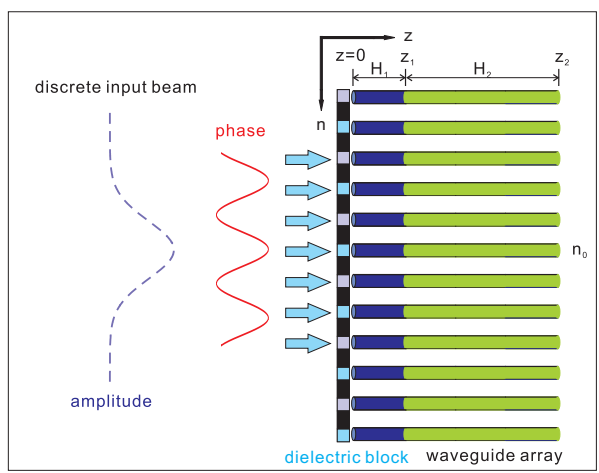}
  \caption{(Color online) Schematic diagram for the elliptical
optical waveguide arrays and the discrete input beam. The light propagates along the axis of waveguide, that is, the $z$
direction. The waveguide array is labeled by $n$ ($n = 1, 2,
...\,, N$). The amplitude of the input beam has a Gaussian distribution. The phase differences between different waveguides are finite, which can be obtained by putting a dielectric block just in front of the array. EOWA.eps}
  \label{fig:EOWA}
  \end{figure}

\begin{figure}[!h t]
  \centering
  \includegraphics[width=0.5 \textwidth]{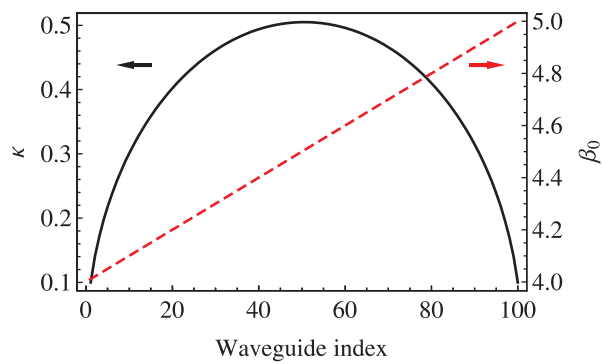}
  \caption{(Color online) In the elliptical optical waveguide arrays, the elliptical potential profile is formed by the circular couplings (solid line) and a linear onsite propagation constant (dashed line). Coup.eps}
  \label{fig:COUP}
  \end{figure}

The EOWA consists of $N=100$ waveguides as shown schematically in figure \ref{fig:EOWA}. This array is divided into two zones (zone 1: $0 \leq z \leq z_1$ and zone 2: $z_1 < z \leq z_2$) along the longitudinal $z$-axis direction, where there are two elliptical potential profiles $H_1$ and $H_2$, respectively. As shown in figure \ref{fig:COUP}, each elliptical potential profile is formed by the circular couplings between adjacent waveguides ($\kappa$, solid line) and a linearly graded onsite propagation constant ($\beta_{0}$, dashed line). The switching between BO and DO can be achieved by ramping up the gradient of the propagation constant from zone 1 to zone 2. The circular couplings are realized by carefully designing the structure of EOWA. The linear gradient in propagation constants are obtained by taking advantage of the electro-optical effect or thermo-optical effect \cite{OptLett.23.1701.1998, PhysRevLett.83.4752.1999}, and the the gradient strength is adjusted by the external voltage or temperature difference. The size of each waveguide is in the micrometer scale. However, the real physical parameters should be calibrated through experiments. As shown in figure \ref{fig:EOWA}, the intensity of the discrete input beam has a Gaussian distribution and its wave front is not a plane wave. The finite phase difference of excitations between adjacent waveguides is realized by putting a dielectric block just in front of the waveguide array. The light beam propagates along the axis of the waveguide array, that is, the $z$ direction. The waveguide array is labeled by $n$ ($n = 1, 2, ..., N$) in the transverse direction.

As mentioned in the literature \cite{PhD_HTrompeter}, the waveguide arrays are analyzed by the coupled-mode theory \cite{OptLett.23.1701.1998}. The evolutionary equation of modal amplitude $a_n$ in the $n$th waveguide is written as
\begin{equation}\label{eq:modal}
\left(\mathbf{i}\frac{d}{dz}+ V_n \right)a_n(z) + \kappa_{n,n-1}a_{n-1}(z)+  \kappa_{n,n+1}a_{n+1}(z) = 0\,,
\end{equation}
where $V_n = \alpha n/N + V_{0}$ is the onsite propagation constant, $\alpha$ is the gradient of the linear potential, $V_{0}$ is the onset propagation constant, which is set to be $4$ in this study, and $\kappa_{n,n-1}$ and $\kappa_{n,n+1}$ are the coupling constants between nearest-neighbor waveguides. In order to obtain harmonic BO and DO, the values of the coupling constants should be designed appropriately, as described in the following.

Substituting the solution $a_n^m(z)=u_n^m\exp({i\beta_m z})$ into (\ref{eq:modal}), we have
\begin{equation}\label{eq:amplitude}
\beta_m u_n^m = V_n u_n^m + \kappa_{n,n-1}u_{n-1}^m+  \kappa_{n,n+1}u_{n+1}^m\,,
\end{equation}
where $\beta_m$ means the wavenumber of the supermode $m$ and the transverse mode profile is given by a superposition of the mode amplitudes
$u_{n}^{m}$ of the individual waveguides. Equation (\ref{eq:amplitude}) is rewritten in the matrix form
\begin{equation}\label{eq:matrix}
\beta|u\rangle = \textbf{H} |u\rangle\,,
\end{equation}
where the Hamiltonian matrix $\textbf{H}$ is defined as $H_{n,n} =
V_n=\alpha n/N + 4$, $H_{n,n-1}= \kappa_{n,n-1}$, and $H_{n,n+1}= \kappa_{n,n+1}$. In order to achieve the harmonic BO and DO, the eigenvalues of $\textbf{H}$ should be equally spaced. It was shown that the $N+1$ by $N+1$ Kac matrix $K$ has equally spaced eigenvalues $2n-N$ ($n = 0, 1, ..., N$) \cite{Algebr.150.341.1991, Proceedings.503.1994}. The Kac matrix is a tridiagonal matrix with elements $K_{n+1,n}= n$, $K_{n,n+1}= N-n$, and $K_{n, m} = 0$ (otherwise), and is asymmetric (non-Hermitian). The Kac matrix is symmetrized \cite{Proceedings.503.1994} and the super and sub-diagonals of $\textbf{H}$ are defined as \cite{OptLett.29.2752.2004}
\begin{equation}
H_{n,n+1}=H_{n+1,n} = \kappa_{n,n+1}=\kappa_{n+1,n} = \frac{\sqrt{n(N-n)}}{N}\,. \quad (n = 1, 2, ..., N-1)
\end{equation}
The symmetric Hamiltonian matrix $\textbf{H}$ is similar to the Kac matrix in a sense that both their eigenvalues are equally spaced. The column vector $|u\rangle$ and $\beta$ denote the eigenvectors and eigenvalues of $\textbf{H}$, respectively. At a certain gradient $\alpha$, the eigenvalues $\beta$ have equal spacing $\Delta \beta = \sqrt{4+\alpha^2}/N$, and thus the density of states is a constant $1/(N \Delta \beta)= 1/\sqrt{4+\alpha^2}$.

Using the Hamiltonian matrix $\textbf{H}$, (\ref{eq:modal}) is written as a
$z$-dependent equation
\begin{equation}\label{eq:z-eq}
-\mathbf{i}\frac{d}{dz} |u\rangle = \mathbf{H} |u\rangle\,.
\end{equation}
It is analogous to the Schr$\ddot{\rm{o}}$dinger equation in quantum system,
\begin{equation}\label{eq:Seq}
-\mathbf{i}\frac{d}{dt} |\phi\rangle = \textbf{H} |\phi\rangle\,.
\end{equation}
Here $\hbar$ is taken to be unity. Thus the quantities $(\beta, z)$
in optical waveguide arrays corresponds to $(\omega, t)$ in quantum
system. We refer to the functional dependence of $\beta$ on
transverse wavenumber $k$ as the dispersion relation in periodic
optical waveguide arrays. For graded arrays, we can divide the
infinite waveguide arrays into a large number of sub-waveguide
arrays in the transverse direction, each of which can be regarded as
infinite in size. Based on the treatment of graded system, we can obtain the band structure approximately as follows. The solution
satisfies the relation $u_{n+1} = u_{n} \exp(i k)$, where $k$ is the
transverse wavenumber. Substituting this relation into
(\ref{eq:amplitude}), we obtain the approximate pseudo-dispersion relation
\begin{equation}\label{eq:disp}
\beta(n,k,\alpha) \approx \alpha \frac{n}{N} + 4 + \frac{2\sqrt{n(N-n)}}{N} \cos k\,,
\end{equation}
which is a function of waveguide index $n$, wavenumber $k$ and gradient strength $\alpha$.

In the following calculations, we use the model parameters: the waveguide number $N = 100$, the input waveguide number $n_0 = 40$, the input wavenumber $k_0 = 0$, the width of the input beam $\sigma = 5$, the initial gradient strength $\alpha = 1.0$ and the ramped-up gradient strength $\alpha = 4.0$ in zone 1 and zone 2, respectively.

\section{Results}

\subsection{Various oscillations and normal modes in EOWA}

\begin{figure}[htbp]
  \centering
  \includegraphics[width=0.5 \textwidth]{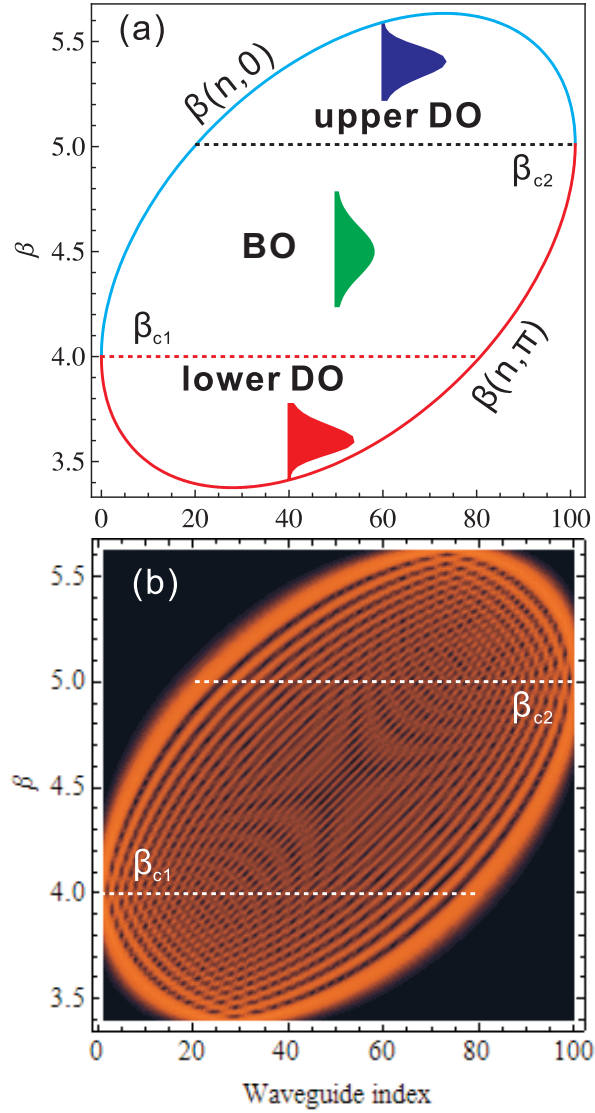}
  \caption{
  (Color online) (a) Phase diagram for the elliptical optical waveguide array with $N=100$ waveguides. The pseudo-dispersion relation lines $\beta(n,0)$ and $\beta(n,\pi)$ serve as the upper and lower boundaries of the band, whose shape is elliptical. Separated by the two critical lines $\beta_{c1}=\beta(1,\pi)$ and $\beta_{c2}=\beta(n,0)$, three regions are formed, which indicate three different kinds of oscillations: lower dipole oscillation, Bloch oscillation and upper dipole oscillation. (b) The contour plots of mode patterns of the eigenvectors in the real space. The shape of the contour plots is similar to the elliptical band. The corresponding mode patterns for the three different oscillations are also separated by the two critical lines $\beta_{c1}$ and $\beta_{c2}$. PDMP.eps}
  \label{fig:PDMP}
  \end{figure}

To analyze the various oscillations in EOWA, we resort to an effective diagrammatic approach with the aid of a phase diagram \cite{PhysRevA.81.033829}. From the pseudo-dispersion relation (\ref{eq:disp}), two curves $\beta(n,0)$ and $\beta(n,\pi)$ serve as the upper and lower boundaries of the phase diagram for EOWA, as shown in figure \ref{fig:PDMP}(a). Separated by two critical lines $\beta_{\rm c1} = \beta(1,\pi)$ and $\beta_{\rm c2} =\beta(N, 0)$, there are three different regions: lower DO region, BO region, and upper DO region. In these three regions, DO at the smaller $n$ side, BO at the middle, and DO at the larger $n$ side take place, respectively. As stated in the previous research \cite{PhysRevA.81.033829}, there exist correspondences between various optical oscillations and the localization of different gradon modes in OWA \cite{PhysRevA.81.033829}. The mode patterns of all the eigenmodes are shown by the contour plots of square moduli of eigenmodes as a function of eigenvalue $\beta$ and waveguide index $n$, as shown in Fig.\ref{fig:PDMP}(b). Separated by $\beta_{\rm c1}$ and $\beta_{\rm c2}$, there are also three different regions, which indicates three different gradon modes corresponding to the three kinds of optical oscillations. If we construct the light beam using components of a specific kind of gradon modes in a particular region, the light beam will undergo a certain type of oscillation, as shown by the three input Gaussian beams in different oscillation regions in figure \ref{fig:PDMP}(a). For a certain input beam, is it possible to undergo DO under certain conditions and BO under other conditions? The answer is yes, we will present it in detail in the following sections.

\subsection{DO-BO transition}

\begin{figure}[htbp]
  \centering
  \includegraphics[width=0.5 \textwidth]{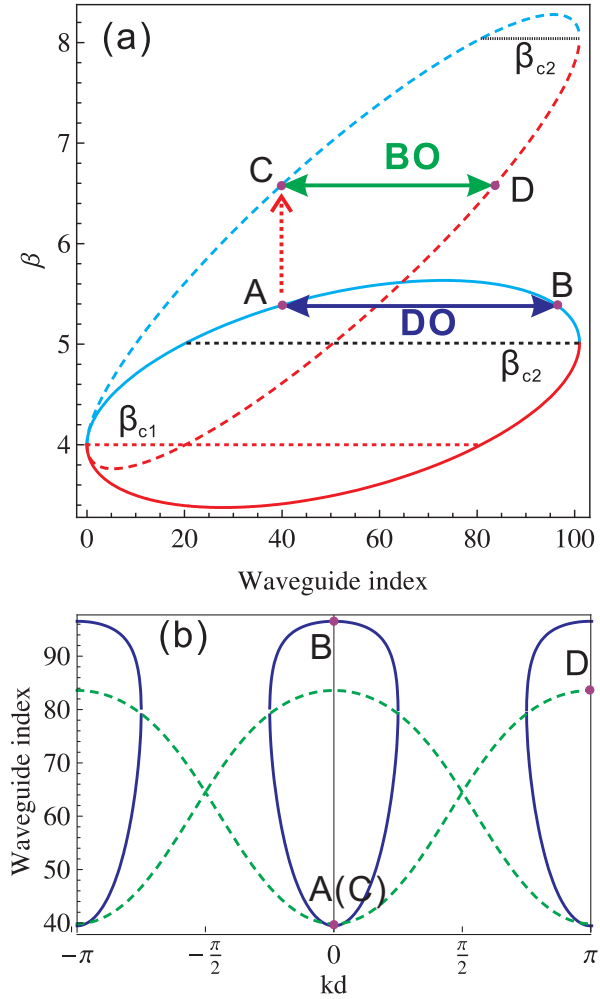}
  \caption{(Color online) (a) A possible DO-BO transition realized by ramping up the value of $\alpha$ from $1.0$ to $4.0$. In the original potential profile, the light beam undergoes DO between points A and B. After several periods, the potential profile is lifted, the light beam is transited from point A to point C and undergoes BO between points C and D. (b) The phase-space orbits for the elliptical optical waveguide arrays when $\alpha = 1.0$ (solid lines) and $\alpha = 4.0$ (dashed lines). The DO-BO transition process A$\rightarrow$ B $\rightarrow$ C $\rightarrow$ D points A and B are also shown. DO-BO.eps}
  \label{fig:DOBO}
  \end{figure}

The switching between DO and BO is realized by ramping up the gradient of the propagation constant. Let us first sketch an example of the DO-BO transition as shown in figure \ref{fig:DOBO}(a). In the range $0\leq z \leq z_1$, the original onsite propagation constant has a linear gradient $\alpha = 1.0$ and the Hamiltonian is $H_1 = \beta(n, k, 1.0)$ (solid lines). This linear gradient together with the circular couplings leads to the long-living DO for an input Gaussian beam ($n_0 = 40$, $k_0 = 0$, $\sigma = 5$). In the original potential profile, the input light beam undergoes DO between points A and B. After two periods, the gradient of propagation constant is changed to be $\alpha = 4.0$ when $z_1\leq z \leq z_2$. As a consequence, the elliptical potential profile is changed to a new one (as the dashed lines showed), the light beam is lifted from point A to point C and undergoes BO between points C and D in the new potential profile. In a word, the DO-BO transition has been realized by varying the gradient strength of propagation constant from $\alpha = 1.0$ to $\alpha = 4.0$. As shown in figure \ref{fig:DOBO}(b), the DO-BO transition process A $\rightarrow$ B $\rightarrow$ C $\rightarrow$ D is also marked on the phase-space orbit. The solid lines are for the original potential profile when $\alpha = 1.0$ and $\beta = 5.38$. Since these curves are closed and their wavenumbers are confined to a certain range, these features indicate the occurrence of DO. The dashed lines are for the new potential profile when $\alpha = 4.0$ and $\beta = 6.58$. The value of wavenumber $kd$ has no limitation along these dashed lines, which indicates the occurrence of BO. The transition takes place between points A and C, which coincide on the phase-space orbit.

\subsection{Field-evolution analysis}

The process of DO-BO transition is investigated through the field-evolution analysis. The analysis is performed with an
input wave function at $z = 0$,
\begin{equation}\label{eq:input}
\psi(0)=\frac{1}{(2\pi\sigma^2)^{1/4}} e^{-\frac{(n - n_0)^2}{4
\sigma^2}}e^{-\mathbf{i}k_0(n - n_0)}\,,
\end{equation}
where $k_0$ is the input transverse wave number. The incoming
field at $z$ ($z < 0$) is $\psi(z)= \psi(0)\exp(\mathbf{i} \beta_0 z)$,
where $\beta_0$ is the propagation constant of individual
homogeneous channel. The intensity profile $|\psi(0)|^2$ has a
discrete Gaussian distribution centered at the $n_0$th
waveguide with spatial width $\sigma$. This input beam is a
discrete Gaussian beam, whose intensity distribution is
schematically shown as the dashed line in figure \ref{fig:EOWA}. The
exponential factor $\exp{[-\mathbf{i}k_0(n - n_0)]}$ denotes the phase
differences between input beams excited on the $n$th and the
$n_0$th waveguides. Other than the Gaussian distribution, the input beam can take other functional forms.

We expand the input wave function in terms of the supermodes
$|u_m\rangle$,
\begin{equation}\label{eq:expansion}
|\psi (0)\rangle = \sum_{m} A_m |u_m\rangle\,,
\end{equation}
where $A_m = \langle u_m|\psi(0)\rangle$ is the constituent
component of the input Gaussian beam. The subsequent
wave function at propagation distance $z$ is given by
\begin{equation}\label{eq:beam}
|\psi (z)\rangle = \sum_{m} A_m e^{\mathbf{i}\beta_m z}|u_m\rangle\,.
\end{equation}
At a certain propagation distance $z$, the wave function in the reciprocal space can be obtained by taking the following Fourier transform
\begin{equation}\label{eq:phik}
|\phi (k,z)\rangle = \mathcal {F}[|\psi (n,z)\rangle]\,.
\end{equation}
By using these wave functions, the mean value of $n$ and $k$ are
obtained as
\begin{equation}\label{eq:mean}
\langle n \rangle = \frac{\langle \psi|n| \psi \rangle}{\langle
\psi|\psi \rangle} \,, \qquad \langle k \rangle = \frac{\langle
\phi|k|\phi\rangle}{\langle \phi|\phi \rangle} \,.
\end{equation}

The field evolution of DO-BO transition is demonstrated by the contour plots of $|\psi(n,z)|^2$ in the real space and $|\phi(k,z)|^2$ in the reciprocal space, as shown in Figs.~\ref{fig:xk}(a) and \ref{fig:xk}(b), respectively. When $0 \leq z \leq z_1$, oscillatory motions take place in both real space and reciprocal space (in a confined range), which indicates the occurrence of DO in this range. When $z_1\leq z \leq z_2$, there is still oscillatory motion in the real space (with a smaller amplitude), which in the reciprocal space, the wavenumber $kd$ is accelerated and can cover the whole range $[-\pi, \pi]$ in the reduced Brillouin zone. These features clearly show the occurrence of DO-BO transition. The period of these harmonic oscillations $Z$ can be obtained from $\psi (z+Z) = \psi (z)$ and (\ref{eq:beam}), where the condition $\exp(\mathbf{i}\Delta \beta Z) = 1$ should be satisfied. Thus the period is derived as $Z=2\pi/\Delta \beta = 2\pi N/\sqrt{4+\alpha^2}$. We need to emphasize that there is almost no leakage of field from the main orbit of field evolution. Both DO and BO can persist for a long propagation distance. This property is important in practical applications of propagating light in optical waveguide arrays.

\subsection{Hamiltonian optics}

\begin{figure}[htbp]
  \centering
  \includegraphics[width=0.8 \textwidth]{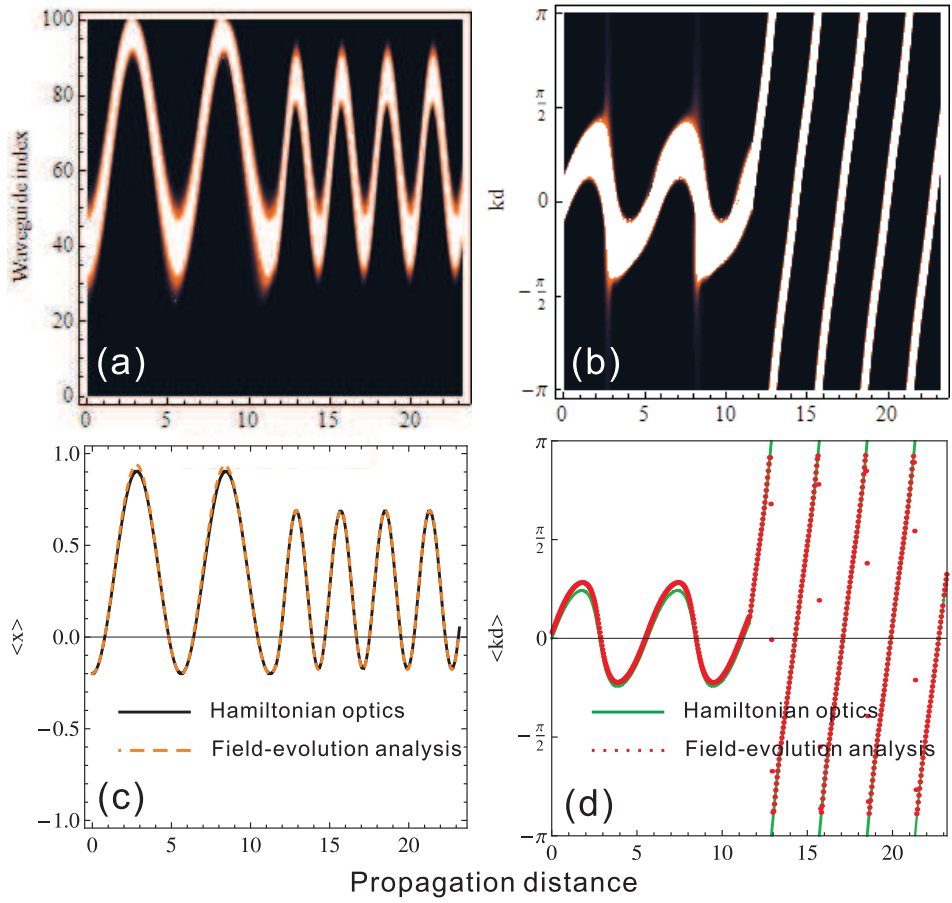}
  \caption{(Color online) Contour plots of field-evolution analysis results for (a) $|\psi(n,z)|^2$ as a function of the waveguide index $n$ and the propagation distance $z$ and (b)$|\phi(k,z)|^2$ as a function of the transverse wavenumber $k$ and the propagation distance $z$. Comparison of Hamiltonian optics results with field-evolution analysis results for (c) $\langle x \rangle$ and (d) $\langle kd\rangle$ in the DO-BO transition. Note that the propagation distance is rescaled by $N/2$. xkComp.eps}
  \label{fig:xk}
  \end{figure}

The results of DO-BO transition obtained through the field-evolution analysis are now confirmed by the Hamiltonian optics approach. From the pseudo-dispersion relation (\ref{eq:disp}), the evolution of the mean position and the mean wavenumber of the beam can be solved by using the equation of motion
\begin{equation}\label{eq:HO}
\frac{dx}{dz} = \frac{\partial \beta(x, k, \alpha)}{\partial k}\,, \qquad
\frac{d k}{dz} = -\frac{\partial \beta(x, k, \alpha)}{\partial x}\,.
\end{equation}
where the auxiliary variable $x = 2n/N-1$ ($-1 < x \leq 1$) is used to replace the discrete waveguide index $n$, and $\beta(x, k, \alpha) = 4+\alpha(1+x)/2+\sqrt{1-x^2}\cos k$. The numerical Hamiltonian optics results of the mean position $\langle x \rangle$ and the mean wavenumber $\langle kd \rangle$ are shown by solid lines in Figs.~\ref{fig:xk}(c) and \ref{fig:xk}(d), respectively. The dashed lines and dots in Figs.~\ref{fig:xk}(c) and \ref{fig:xk}(d) are the mean values of $\langle x \rangle$ and $\langle kd \rangle$ calculated from the field-evolution analysis results. The mean values obtained from two methods match very well with each other.

\section{Discussion and conclusion}

As suggested in Ref.~\cite{OptLett.29.2752.2004}, the experimental realization of the EOWA is similar to the experimental structures GaAs/AlGaAs \cite{PhysRevLett.83.4756.1999}. To obtain the required $\kappa$ values, the spacing between the waveguides was carefully designed based on the evanescent codirectional couplings \cite{A.Yariv}. In the above investigation, we considered the nearest-neighbor couplings between waveguides. This consideration is valid when the field overlapping is significant between neighboring waveguides only. We can also extend this study to more general OWAs with higher-order couplings. As is shown in the reference \cite{OptLett.35.1908.2010}, the second-order coupling in OWA leads to nontrivial BO. It is instructive to consider higher-order couplings and the effects of defects in EOWA. Isolated defects can always disrupt the equally spaced eigenvalues and lead to damping, but Longhi designed localized defects carefully by supersymmetric quantum mechanics and obtained undamped  optical oscillations \cite{PhysRevB.81.195118.2010}. From the basic physics point of view, an intrinsic harmonic system can naturally help us to access the competing damping mechanisms, like defects and disorder or even nonlinearity. Thus in the future work, we will investigate the nature of the harmonic oscillation against these effects.

In summary, we have studied the dipole oscillation and Bloch oscillation in the elliptical optical waveguide arrays, which possess a linear gradient in the onsite propagation constants and circular couplings between neighboring waveguides. The couplings between neighboring waveguides are designed according to the tridiagonal elements of a Kac matrix, which is to achieve equally spaced eigenvalues. As a consequence, both DO and BO are long-living, which is important to optical steering applications. We also proposed to realize the DO-BO transition or vice versa by ramping up the gradient of propagation constants. The spatial evolution of the DO-BO transition is demonstrated through the field-evolution analysis and confirmed by the Hamiltonian optics approach. The long-living DO and BO and their switching have viable applications in transmitting images, switching and routing of optical signals.

\section*{Acknowledgments}

This work was supported by RGC General Research Fund of the
Hong Kong SAR Government. We thank Prof. K. Yakubo for careful
reading of the manuscript and for many useful discussion and
helpful suggestions.

\newpage

%\section*{REFERENCES}
% Bibliography
%\bibliographystyle{phaip}
%\pagestyle{plain}
%\bibliography{E:/Research/Biblibrary/biblib}

\begin{thebibliography}{99}

\bibitem{PhysRevA.63.063410.2001}
Schirmer S G, Fu H, and Solomon A I 2001 Complete controllability of quantum systems
\newblock Phys. Rev. A {\bf 63}, 063410

\bibitem{IEEE.49.745.2004}
Mirrahimi M and Rouchon P 2004 Controllability of Quantum Harmonic Oscillators
\newblock IEEE Trans. Automat. Contr. {\bf 49}, 745-747

\bibitem{OptLett.29.2752.2004}
Gordon R 2004 Harmonic oscillation in a spatially finite array waveguide
\newblock Opt. Lett. {\bf 29}, 2752-4754

\bibitem{IEEE.71.208.1983}
Friesem A A, Levy U, and Silverberg Y 1983 Digital and analog optical broad-band transmission
\newblock Proc. IEEE {\bf 71}, 208

\bibitem{PhysRevLett.103.033902.2009}
Verslegers L, Catrysse P B, Yu Z, and Fan S 2009 Deep-subwavelength focusing and steering of light in an aperiodic metallic waveguide array
\newblock Phys. Rev. Lett. {\bf 103}, 033902

\bibitem{PhysRevLett.96.023901.2006}
Trompeter H, Pertsch T, Lederer F, Michaelis D, Streppel U, and Brauer A  2006 Visual Observation of Zener Tunneling
\newblock Phys. Rev. Lett. {\bf 96}, 023901

\bibitem{PhysRevLett.96.053903.2006}
Trompeter H,  Krolikowski W, Neshev D N, Desyatnikov A S, Sukhorukov A A, Kivshar Y S, Pertsch T, Peschel U and Lederer F
2006 Bloch Oscillations and Zener Tunneling in Two-dimensional Photonic Lattices
\newblock Phys. Rev. Lett. {\bf 96}, 053903

\bibitem{PhysRevLett.102.076802.2009}
Dreisow F, Szameit A, Heinrich M, Pertsch T, Nolte S, Tunnermann A, and Longhi S  2009 Bloch-Zener Oscillations in Binary Superlattices
\newblock Phys. Rev. Lett. {\bf 102}, 076802

\bibitem{PhysRevLett.83.4752.1999}
Pertsch T, Dannberg P, Elflein W, Brauer A, and Lederer F  1999 Optical Bloch oscillations in temperature tuned waveguide arrays
\newblock Phys. Rev. Lett. {\bf 83}, 4752-4755

\bibitem{PhysRevLett.83.4756.1999}
Morandotti R, Peschel U, Aitchison J S, Eisenberg H S, and Silberberg Y 1999 Experimental observation of linear and nonlinear optical Bloch oscillations
\newblock Phys. Rev. Lett. {\bf 83}, 4756-4759

\bibitem{LaserPhoton.Rev.3.243.2009}
Longhi S 2009 Quantum-optical analogies using photonic structures
\newblock Laser and Photon. Rev. {\bf 3}, 243-261

\bibitem{Nature.424.817.2003}
Christodoulides D N, Lederer F, and Silberberg Y 2003 Discretizing light behaviour in linear and nonlinear waveguide lattices
\newblock Nature {\bf 424}, 817-823

\bibitem{PhysRep.463.1.2008}
Lederer F, Stegemanb G I, Christodoulides D N, Assanto G, Segev M, and Silberberg Y 2008 Discrete solitons in optics
\newblock Phys. Rep. {\bf 463}, 1-126

\bibitem{NewJPhys.6.2.2004}
Hartmann T, Keck F, Korsch,H J and Mossmann S 2004 Dynamics of Bloch oscillations
\newblock New J. Phys. {\bf 6}, 2-25

\bibitem{OptLett.23.1701.1998}
Peschel U, Pertsch T, and Lederer F 1998 Optical Bloch oscillations in waveguide arrays
\newblock Opt. Lett. {\bf 23}, 1701-1703

\bibitem{PhysRevLett91.263902.2003}
Sapienza R, Costantino P, Wiersma D, Ghulinyan M, Oton C J, Pavesi L 2003 Optical analogue of electronic Bloch oscillations
\newblock Phys. Rev. Lett. {\bf 91}, 263902

\bibitem{NewJPhys.5.112.2003}
Menotti C, Smerzi A, and Trombettoni A 2003 Superfluid dynamics of a Bose-Einstein condensate in a periodic potential
\newblock New J. Phys. {\bf 5}, 112.1-112.20

\bibitem{PhysRevA.81.033829}
Zheng M J, Xiao J J, and Yu K W 2010 Controllable optical Bloch oscillation in planar graded optical waveguide arrays
\newblock Phys. Rev. A {\bf 81}, 033829

\bibitem{JOptSocAmB.27.1299.2010}
Zheng M J, Chan Y S, and Yu K W 2010 Steering between Bloch oscillation and dipole oscillation in parabolic optical waveguide arrays
\newblock J. Opt. Soc. Am. B {\bf 27}, 1299-1304

\bibitem{OptLett.29.2485.2004}
Efremidis N K and Christodoulides D N 2004 Bloch oscillations in optical dissipative lattices
\newblock Opt. Lett. {\bf 29}, 2485-2487

\bibitem{PhysRevB.81.195118.2010}
Longhi S  Bloch oscillations in tight-binding lattices with defects 2010
\newblock Phys. Rev. B {\bf 81}, 195118

\bibitem{Algebr.150.341.1991}
Taussky O  and Todd J 1991 Another look at a matrix of Mark Kac
\newblock Linear Algebra Appl. {\bf 150}, 341-360

\bibitem{Proceedings.503.1994}
Edelman A and Kostlan E 1994 The road from Kac's matrix to Kac's random polynomials
\newblock Proceedings of the Fifth SIAM on Applied Linear Algebra J G. Lewis, Ed, Philadelphia, 503

\bibitem{PhD_HTrompeter}
Trompeter H 2006
\newblock {\em Discrete optics in inhomogeneous waveguide arrays},
\newblock {PhD} dissertation, Friedrich-Schiller Universitat Jena

\bibitem{A.Yariv}
Yariv A 1997
\newblock {\em Optical Electronics in Modern Communications},
\newblock Oxford University, New York, 5th edition

\bibitem{OptLett.35.1908.2010}
Wang G, Huang J P, and Yu K W 2010 Nontrivial Bloch oscillations in waveguide arrays with second-order coupling
\newblock Opt. Lett. {\bf 35}, 1908-1910

\end{thebibliography}

\clearpage

\section*{List of Figure Captions}

\noindent Fig. 1 (Color online) Schematic diagram for the elliptical
optical waveguide arrays and the discrete input beam. The light
propagates along the axis of waveguide, that is, the $z$
direction. The waveguide array is labeled by $n$ ($n = 1, 2,
.., N$). The amplitude of the input beam has a Gaussian distribution. The phase differences between different waveguides are finite, which can be obtained by putting a dielectric block just in front of the array. EOWA.eps

\noindent Fig. 2 (Color online) In the elliptical optical waveguide arrays, the elliptical potential profile is formed by the circular couplings (solid line) and a linear onsite propagation constant (dashed line). Coup.eps

\noindent Fig. 3 (Color online) (a) Phase diagram for the elliptical optical waveguide array with $N=100$ waveguides. The pseudo-dispersion relation lines $\beta(n,0)$ and $\beta(n,\pi)$ serve as the upper and lower boundaries of the band, whose shape is elliptical. Separated by the two critical lines $\beta_{c1}=\beta(1,\pi)$ and $\beta_{c2}=\beta(n,0)$, three regions are formed, which indicate three different kinds of oscillations: lower dipole oscillation, Bloch oscillation and upper dipole oscillation. (b) The contour plots of mode patterns of the eigenvectors in the real space. The shape of the contour plots is similar to the elliptical band. The corresponding mode patterns for the three different oscillations are also separated by the two critical lines $\beta_{c1}$ and $\beta_{c2}$. PDMP.eps

\noindent Fig. 4 (Color online) (a) A possible DO-BO transition realized by ramping up the value of $\alpha$ from $1.0$ to $4.0$. In the original potential profile, the light beam undergoes DO between points A and B. After several periods, the potential profile is lifted, the light beam is transited from point A to point C and undergoes BO between points C and D. (b) The phase-space orbits for the elliptical optical waveguide arrays when $\alpha = 1.0$ (solid lines) and $\alpha = 4.0$ (dashed lines). The DO-BO transition process A$\rightarrow$ B $\rightarrow$ C $\rightarrow$ D points A and B are also shown. DO-BO.eps

\noindent Fig. 5 (Color online) Contour plots of field-evolution analysis results for (a) $|\psi(n,z)|^2$ as a function of the waveguide index $n$ and the propagation distance $z$ and (b)$|\phi(k,z)|^2$ as a function of the transverse wavenumber $k$ and the propagation distance $z$. Comparison of Hamiltonian optics results with field-evolution analysis results for (c) $\langle x \rangle$ and (d) $\langle kd\rangle$ in the DO-BO transition. Note that the propagation distance is rescaled by $N/2$. xkComp.eps

%\listoffigures

\clearpage

\singlespace

%% sample sizing command; other sizing commands (and graphics packages) may be used as well
\newpage

%\newpage

\newpage

  \newpage

\newpage

\end{document}